\documentclass[journal]{IEEEtran}

\usepackage{amsmath,amssymb,pstricks,color,dsfont}
\usepackage{savesym}
\savesymbol{AND}
\usepackage{graphicx}
\usepackage{psfrag,graphicx,epsfig}
\usepackage{algorithm,algorithmic,xspace}
\usepackage{subfig}
\usepackage{float}
\usepackage{graphicx}
\usepackage{epstopdf}
\usepackage{url}
\newcommand{\lbd}{{\lambda}}

\newcommand{\nnum}{\nonumber}

\newcommand{\EQ}{\begin{eqnarray}}
\newcommand{\EN}{\end{eqnarray}}
\newcommand{\EQQ}{\begin{eqnarray*}}
\newcommand{\ENN}{\end{eqnarray*}}

\newcommand{\bremark}{\begin{remark} \begin{rm} }
\newcommand{\eremark}{ \end{rm} \rule{1mm}{2mm}
\end{remark} }
\newcommand{\btheorem}{\begin{theorem} \begin{rm} }
\newcommand{\etheorem}{ \end{rm} \rule{1mm}{2mm}
\end{theorem} }
\newcommand{\blemma}{\begin{lemma} \begin{rm} }
\newcommand{\elemma}{ \end{rm} \rule{1mm}{2mm}
\end{lemma} }
\newcommand{\bcorollary}{\begin{corollary} \begin{rm} }
\newcommand{\ecorollary}{ \end{rm} \rule{1mm}{2mm}
\end{corollary} }
\newcommand{\bdefinition}{\begin{definition}\begin{rm} }
\newcommand{\edefinition}{ \end{rm} \rule{1mm}{2mm}
\end{definition} }
\newcommand{\bproposition}{\begin{proposition} \begin{rm} }
\newcommand{\eproposition}{ \end{rm} \rule{1mm}{2mm}
\end{proposition} }
\newcommand{\bexample}{\begin{example} \begin{rm} }
\newcommand{\eexample}{ \end{rm} \rule{1mm}{2mm}
\end{example} }
\newcommand{\basm}{\begin{assumption} \begin{rm}}
\newcommand{\easm}{\end{rm} 
\end{assumption}}

\newcommand{\real}{\mathds{R}}

\newcommand{\EE}{{\mathcal{E}}}

\newcommand{\GG}{{\mathcal{G}}}

\newcommand{\KK}{{\mathcal{K}}}

\newcommand{\LL}{{\mathcal{L}}}
\newcommand{\NN}{{\mathcal{N}}}

\newcommand{\VV}{{\mathcal{V}}}

\newcommand{\diag}[1]{\operatorname{diag}(#1)}

\newtheorem{theorem}{\bf Theorem}[section]
\newtheorem{lemma}{\bf Lemma}[section]
\newtheorem{definition}{\bf Definition}[section]
\newtheorem{remark}{\bf Remark}[section]
\newtheorem{corollary}{\bf Corollary}[section]
\newtheorem{assumption}{\bf Assumption}[section]

\newcommand\oprocendsymbol{\hbox{$\blacksquare$}}
\newcommand\oprocend{\relax\ifmmode\else\unskip\hfill\fi\oprocendsymbol}

\date{}

\begin{document}

\title{Distributed Robust Adaptive Frequency Control of Power Systems with Dynamic Loads}
\author{
Hunmin Kim, Minghui Zhu, and Jianming Lian
\thanks{This work was supported in part by the Technical area of Sensing and Measurements within Grid Modernization Initiative funded by the U.S. DOE. \textit{(Corresponding author: Jianming Lian.)}}
\thanks{H. Kim is with
the Department of Mechanical Science and Engineering, University of Illinois at Urbana-Champaign, Urbana, IL 61801 USA (email: hunminkim3@gmail.com).}
\thanks{M. Zhu is with the School of Electrical Engineering and Computer Science, Pennsylvania State University, University Park, PA 16802 USA (email: muz16@psu.edu).}
\thanks{
J. Lian is with the Optimization and Control Group, Pacific Northwest National Laboratory, Richland, WA 99352 USA (email: jianming.lian@pnnl.gov).}
}

\maketitle

\begin{abstract}
This paper investigates the frequency control of multi-machine power systems subject to uncertain and dynamic net loads. We propose distributed internal model controllers that coordinate synchronous generators and demand response to tackle the unpredictable nature of net loads. Frequency stability is formally guaranteed via Lyapunov analysis. Numerical simulations on the IEEE 68-bus test system demonstrate the effectiveness of the controllers.
\end{abstract}

\section{Introduction}
The power grid is modernized into the smart grid. The wide deployment of advanced information and communications technologies facilitates real-time pricing and demand response. In addition, centralized generating facilities are giving way to small distributed energy resources; e.g., photovoltaic systems, fuel cells, storage and electric vehicles. Moreover, renewable energy; e.g., wind, solar and wave energy, has been increasingly adopted due to its cleanness and profitability. 
Worldwide 144 countries now have their political targets for increase in renewable generation, and especially the European Union sets a goal of 20\% share of renewable energy by 2020~\cite{ren212014global}.
However, renewable generation is hard to predict, and an increasing proliferation of renewable energy imposes significant challenges to the operation and management of the power grid. It becomes imperative to maintain grid stability and reliability despite the disturbances from renewable generation.

\textbf{Literature review.} There has been rich literature on control of the power grid under a variety of external disturbances.
Representative techniques include Riccati equation~\cite{lim1996robust,ray1999new}, $H_2/H_{\infty}$ control~\cite{bevrani2005bilateral,shayeghi2008robust}, and LMIs~\cite{marinovici2013distributed,siljak2002robust}.
This set of papers focuses on disturbance attenuation where the impacts of external disturbances are reduced but not completely eliminated and so only practical stability can be achieved. Disturbance rejection instead aims to completely eliminate external disturbances and recover perfect stability. The literature on disturbance rejection in power systems is limited. A recent work~\cite{trip2016internal} develops distributed internal model controllers to ensure optimal frequency synchronization despite uncertain and time-varying loads. There are a couple of distinctions from this paper. First, work in~\cite{trip2016internal} assumes that each load consists of a set of sinusoidal functions and their frequencies are known. The assumption of known load frequencies is relaxed in the current paper by the technique of adaptive internal model. Second, work in~\cite{trip2016internal} studies frequency synchronization (primary control) and this paper instead aims to bring grid frequencies back to a nominal value (secondary control).
Besides, distributed control of power systems has been receiving substantial attention (see an incomplete list~\cite{simpson2013synchronization,xin2011self,zhang2015real,zhao2014design}). This set of papers does not consider time-varying external disturbances.

\textbf{Contribution statement.} We study the frequency control of multi-machine power systems with uncertain and dynamic net loads.
Under the assumption that each net load consists of a set of sinusoidal functions, two cases are studied:
(1) \emph{robust adaptive frequency control}, where the frequencies of net loads are unknown; (2) \emph{robust frequency control}, where the frequencies of net loads are known but dynamic systems are more general.
We design distributed internal model controllers, which coordinate synchronous generators and demand response to address the two cases. Frequency stability is guaranteed via Lyapunov analysis. It is the first time to study distributed adaptive internal model control to handle external disturbances with unknown frequencies. As a byproduct, we develop a distributed constrained small-gain theorem. Simulations verify the performance of the distributed controllers on the IEEE 68-bus power system.
It is assumed in this paper as in most of the literature that there is sufficient reactive power support to ensure voltage stability during frequency control.
Preliminary results were published in~\cite{HK-MZ:ACC15} where \emph{robust adaptive frequency control} was not studied and proofs were omitted.

\textbf{Notations.}
Denote $\|x\|_{[t_1,t_2]} \triangleq \sup_{t_1 \leq t \leq t_2} \|x(t)\|$.
Let $|\mathcal{S}|$ be the cardinality of a set $\mathcal{S}$.
Matrix $I_n$ denotes the $n \times n$ identity matrix. Let $\diag{A_1,\cdots,A_n}$ denote a block matrix having $A_1$ to $A_n$ as main diagonal blocks. 
For $v \in {\real}^n$, $sgn(v) \in \{-1,0,1\}^n$ is a sign function, where $sgn(v_i)=-1$ if $v_i<0$, $sgn(v_i)=0$ if $v_i=0$, and $sgn(v_i)=1$ if $v_i>0$. Norm $\|\cdot\|_F$ denotes Frobenius norm.

\section{System Model}\label{sec:problem-formulation}

Table~\ref{tab:1} summarizes the notations in the model. We use $\Delta$ to represent deviations from nominal values; e.g., $\Delta w(t) = w(t) - w^*$, where $w^*$ is the nominal value of $w(t)$. 

\subsection{Power network model}\label{sec:power-model}
The power network is described by the undirected graph $({\VV},\EE)$ where ${\VV} \triangleq \{1,\cdots,N\}$ denotes the set of buses and $\EE \subseteq {\VV} \times {\VV}$ denotes the set of transmission lines between the buses. The set ${\NN}_i$ denotes the set of neighboring buses of $i \in {\VV}$; i.e., ${{\NN}}_i \triangleq \{j \in {\VV} \setminus \{i\} | (i,j) \in \EE\}$.
Each bus is either a generator bus $i \in {\GG}$, or a load bus $i \in {\LL}$ where ${\GG}$ and ${\LL}$ denote the sets of corresponding buses, respectively. Each bus $i$ is associated with a local control authority.
\begin{table}[t]
 \def\arraystretch{1.00} \centering \normalsize
\caption{System variables and parameters}
\begin{tabular}{clcl}
\multicolumn{4}{l}{{System variables}}\\
\hline 
$w$ &  angular frequency & $\theta$ & phase angle \\
$P_M$ &  mechanical power & $P_{ij}$ & power flow \\
$P_v$ &  steam valve position & $P_{ref}$  & reference power\\
$P_C$ &  controllable load& $P_{L}$ & net load\\
\multicolumn{4}{l}{{System parameters}}\\
\hline
$D$ &  damping constant & $m$ & angular momentum\\
$T_{CH}$ & charging time const. & $t_{ij}$ & tie-line stiffness\\
$T_G$ & governor time const.  & $R$ & feedback loop gain\\
$K_m$ & turbine gain & $K_e$ & governor gain\\ 
\end{tabular}
\vspace{-0.5cm}
\label{tab:1}
\end{table}

\subsection{Load model}
An electrical load can be divided to a controllable load and an uncontrollable load~\cite{alvarado1999stability,Alvarado.Meng.ea:01,chiu2013energy}.
Controllable load $\Delta P_{C_i}(t)$ is governed by the demand response~\cite{Alvarado.Meng.ea:01}:
\begin{align}
\Delta \dot{P}_{C_i}(t) &= b_i + c_i \Delta P_{C_i}(t) - \Delta \lbd_i(t)
\label{e6.04}
\end{align}
where $b_i + c_i \Delta P_{C_i}(t)$ is marginal benefit with $c_i<0$ and real-time electricity price $\Delta \lbd_i(t)$ is used as the input for $i \in \LL$.

For $i\in \LL$, net load $\Delta P_{L_i}(t)$ represents the difference between the uncontrollable load and renewable generation. For $i\in \GG$, net load $\Delta P_{L_i}(t)$ represents renewable generation.
Notice that uncontrollable loads and renewable generation are hard to predict.
We regard net loads $\Delta P_{L_i}(t)$ as external disturbances to the power system.
According to the spectral decompositions of wind generation~\cite{milan2013turbulent,van1957power} and load pattern~\cite{aguirre2008dynamical}, we approximate each net load as the sum of a finite number of distinct sinusoidal functions as in~\cite{trip2016internal}.
In addition, any periodic function can be represented by a Fourier series.
If a function is continuous, absolutely integrable and its derivative is absolutely integrable, then its Fourier series converges uniformly to the function (Theorem on p.86 in~\cite{tolstov2012fourier}).
As output regulation~\cite{francis1975internal,isidori1990output}, the following marginally stable exosystem is used to generate $\Delta P_{L_i}(t)$:
\begin{align}
\dot{\chi}_i(t) = \Phi_i(\rho_i) \chi_i(t), \ \Delta P_{L_i}(t)& = \Psi_i \chi_i(t)
\label{e21}
\end{align}
where $\chi_i(t)= [\chi_{i,1}(t), \dot{\chi}_{i,1}(t), {\small \cdots}, \chi_{i,\ell_i}(t),\dot{\chi}_{i,\ell_i}(t)]^T \in \real^{2\ell_i}$,
\begin{align*}
\Phi_i(\rho_i) \triangleq \diag{\Phi_{i,1}, \cdots, \Phi_{i,\ell_i}}, \ \Phi_{i,l} \triangleq
\left[
\begin{array}{cc}
0&1\\
-(\rho_{i,l})^2&0\\
\end{array}
\right].
\end{align*}
Each state $\chi_{i,l}(t)$ is a sinusoidal function with frequency $\rho_{i,l}$. The output $\Delta P_{L_i}(t)$ is then a linear combination of sinusoidal functions with frequencies $\rho_i=\{\rho_{i,1}, \cdots, \rho_{i,\ell_i}\}$.
\begin{assumption}
The pair $(\Psi_{i},\Phi_i(\rho_i))$ is observable.
\label{asmp3}
\end{assumption}

\subsection{Dynamic model of the generator buses}\label{sec:gen0}
Consider the synchronous power generator from~\cite{Wood:1996}:
\begin{align}
\Delta\dot{\theta_i}(t) &= \Delta w_i(t)\nnum\\
\Delta\dot{w_i}(t) &= -\frac{1}{m_i}\big((D_{G_i}+D_{L_i}) \Delta w_i(t) + \sum_{j\in{{\NN}}_i} \Delta P_{ij}(t) \nnum\\
&+\Delta P_{L_i}(t)- \Delta P_{M_i}(t)\big)\nnum\\
\Delta\dot{P}_{M_i}(t) &= -\frac{1}{T_{{CH}_i}}\big(\Delta P_{M_i}(t) - K_{m_i}\Delta P_{v_i}(t)\big)\nnum\\
\Delta\dot{P}_{v_i}(t) &= -\frac{1}{T_{G_i}}\big(\Delta P_{v_i}(t) + \frac{K_{e_i}}{R_i} \Delta w_i(t) - \Delta P_{{ref}_i}(t)\big)
\label{e24}
\end{align}
for $i \in {\GG}$, where $D_{G_i}$ is mechanical damping constant, and $D_{L_i}$ is load damping constant corresponding to net load (renewable generation).
Power flow $\Delta P_{ij}(t)$ is described by $\Delta P_{ij}(t) = t_{ij}(\Delta \theta_i(t) - \Delta \theta_j(t))$.
The first equation in~\eqref{e24} indicates the evolution of phase angle $\Delta \theta_i(t)$.
The second equation is referred to as swing dynamics, indicating frequency fluctuations due to power imbalances.
The third and fourth equations represent turbine governor dynamics with reference input $\Delta P_{ref_i}(t)$.

\subsection{Dynamic model of the load buses}
A load bus $i\in {\LL}$ can be modeled by the following phase angle dynamics and swing dynamics~\cite{ilic2010modeling}, where $m_i$ is the effective moment of a postulated load model:
\begin{align}
&\Delta\dot{\theta_i}(t) = \Delta w_i(t)\nnum\\
&\Delta\dot{w_i}(t) = -\frac{1}{m_i}\big(D_{L_i} \Delta w_i(t) + \sum_{j \in \NN_i}\Delta P_{ij}(t) + \Delta P_{{C}_i} \nnum\\
&\quad \quad \quad +\Delta P_{{L}_i}(t)\big)
\label{eq_load}
\end{align}
where $D_{L_i}$ is load damping constant corresponding to net load which is the difference between the uncontrollable load and renewable generation.

\subsection{Outputs and inputs}\label{sec:outcon}
Control authority $i$ can access $\Delta y_i(t)=[\Delta w_i(t)$, $\Delta P_{M_i}(t),  \Delta P_{v_i}(t),\Delta P_{\NN_i}(t)]^T$ for $i \in {\GG}$, and $\Delta y_i(t)=[\Delta w_i(t), \Delta P_{C_i}(t),  \Delta P_{\NN_i}(t)]^T$ for $i \in {\LL}$, where $\Delta P_{\NN_i}(t) \triangleq \sum_{j \in {{\NN}}_i}\Delta P_{ij}(t)$. Local inputs are $u_i(t) = \Delta P_{ref_i}(t)$ for $i \in {\GG}$, and $u_i(t) = \Delta \lbd_i(t)$ for $i \in {\LL}$.

\section{Problem statement}\label{sec:controller}
In this paper, we investigate the frequency control; i.e., controlling $\Delta w_i(t)$ to zero, and discuss two cases where the frequencies $\rho_i$ of net loads are known or unknown.
We drop $\Delta$ in the rest of the paper for notational simplicity. Also, we use $D_i=D_{G_i}+D_{L_i}$ for $i \in \GG$ and $D_i=D_{L_i}$ for $i \in \LL$. 

\subsection{Case 1: Robust frequency control.}\label{sec:formulation}
To stabilize the frequencies, each generator bus $i \in \GG$ aims to approach the following manifolds:
\begin{align}
w_i^* &= 0, \ P_{M_i}^*(t) = P_{L_i}(t)+P_{\NN_i}(t),\nnum\\
P_{v_i}^*(t) &=\frac{T_{{CH}_i}}{K_{m_i}} \dot{P}_{M_i}^*(t) + \frac{1}{K_{m_i}}P_{M_i}^*(t)\nnum\\
P_{{ref}_i}^*(t) &= T_{G_i} \dot{P}_{v_i}^*(t) + P_{v_i}^*(t)
\label{e19.03}
\end{align}
which can be easily derived from system~\eqref{e24}.
Similarly, each load bus $i \in \LL$ is expected to stay on the following manifolds:
\begin{align}
w_i^* &= 0, \ P_{C_i}^*(t) = -P_{L_i}(t)-P_{\NN_i}(t),\nnum\\
\lbd_i^*(t) &= b_i + c_i P_{C_i}^*(t) - \dot{P}_{C_i}^*(t)
\label{e19.04}
\end{align}
which can be derived from systems~\eqref{e6.04} and~\eqref{eq_load}.
Superscript $*$ denotes the manifold; e.g., $\lambda_i^*(t)$ denotes the manifold of $\Delta \lambda_i(t)$.
We desire to design a distributed controller which steers the system states and inputs to their manifolds~\eqref{e19.03} and~\eqref{e19.04}.
In this case, 
control authority $i$ knows $\rho_i$, $\Psi_i$ and $\Phi_i(\rho_i)$ but is unaware of initial state $\chi_i(0)$ and state $\chi_i(t)$ of exosystem~\eqref{e21}. That is, control authority $i$ knows the number of sinusoidal signals, and their frequencies, but not their phase shifts and magnitudes.
\begin{assumption}
All the frequencies $\rho_{i,1}, \cdots, \rho_{i,\ell_i}$ in~\eqref{e21} of net load $P_{L_i}(t)$ are known to local control authority $i$.
\label{atsm1}
\end{assumption}

\subsection{Case 2: Robust adaptive frequency control.}\label{sec:formulation2}
There are a couple of distinctions from \emph{Case 1}.
First, control authority $i$ is unaware of frequencies $\rho_i$ and Assumption~\ref{atsm1} is weakened into the following one:
\begin{assumption}
Control authority $i$ knows the value $\ell_i$ and an upper bound $\rho_{\max} \geq \max_{i,l}\rho_{i,l}$.
\label{atsm3}
\end{assumption}
Secondly, we use the following simplified synchronous generator model~\cite{Wood:1996}:
\begin{align}
\dot{\theta_i}(t) &= w_i(t)\nnum\\
\dot{w_i}(t) &= -\frac{1}{m_i}\big(D_i w_i(t) + \sum_{j\in{{\NN}}_i} P_{ij}(t) + P_{L_i}(t)- P_{M_i}(t) \big)
\label{e24_2}
\end{align}
and simplified demand response model; i.e., local control authority $i$ controls $P_{C_i}(t)$ directly.
Remark~\ref{rem:sec5} in Section~\ref{sec:stab2} discusses why the simplified models are needed for Case 2.
For this case, the corresponding manifolds are
\begin{align}
&w_i^* = 0, \ P_{M_i}^*(t) = P_{L_i}(t)+P_{\NN_i}(t), \ i\in \GG\nnum\\
&w_i^* = 0, \ P_{C_i}^*(t) = -P_{L_i}(t)-P_{\NN_i}(t), \ i\in \LL.
\label{e19.03L}
\end{align}

\section{Controller synthesis for robust frequency control}\label{sec:sol1}
In this section, we present a solution of the \emph{robust frequency control} described in Section~\ref{sec:formulation}.

\subsection{Local internal models}\label{sec:internal}
Net loads $P_{L_i}(t)$ cannot be measured and thus manifolds~\eqref{e19.03} and~\eqref{e19.04} cannot be used for feedback control. We adopt the methodology of internal models to tackle this challenge~\cite{isidori1990output,francis1976internal}.
Recall that $P_{L_i}(t)$ is the output of exosystem~\eqref{e21}. Hence, for $i\in \GG$, the second equation in~\eqref{e19.03} can be written as:
\begin{align}
P_{M_i}^*(t)-P_{\NN_i}(t)= P_{L_i}(t)= \Psi_i \chi_i(t).
\label{eint}
\end{align}
Under Assumption~\ref{asmp3}, for any controllable pair $(M_{i}, N_{i})$ with $M_{i} \in \real^{2\ell_i \times 2\ell_i}$ being Hurwitz and $N_i \in \real^{2\ell_i}$, there exists a non-singular matrix $T_{i}(\rho_i) \in \real^{2\ell_i \times 2\ell_i}$ as the unique solution of the following Sylvester equation~\cite{bhatia1997and}:
\begin{align}
T_{i}(\rho_i) \Phi_i(\rho_i) - M_{i}T_{i}(\rho_i) &= N_{i}\Psi_{i}.
\label{e111}
\end{align}
With $\vartheta_{i}(t) \triangleq T_{i}(\rho_i) \chi_{i}(t)$, \eqref{eint} becomes
\begin{align*}
P_{M_i}^*(t)-P_{\NN_i}(t)= \Psi_{i}\chi_{i}(t)=\Psi_{i} T_{i}^{-1}(\rho_i)\vartheta_{i}(t).
\end{align*}
Now consider a local internal model candidate:
\begin{align}
\dot{\eta}_{i}(t) &= M_{i} \eta_{i}(t) + N_{i} (P_{M_i}(t)-P_{\NN_i}(t))
\label{e26}
\end{align}
where $\eta_i(t) \in \real^{2\ell_i}$.
Internal model~\eqref{e26} behaves as an estimator and its states $\eta_i(t)$ are expected to asymptotically track unmeasurable exosystem states $\vartheta_i(t)$. The manifolds of $\eta_{i}(t)$ are $\eta_i^*(t)=\vartheta_i(t)$ in this case.
It is expected to stabilize the dynamics of error $\eta_i(t)-\vartheta_i(t)$.
According to the certainty equivalence principle~\cite{whittle1986risk}, internal model states $\eta_i(t)$ are used to replace $\vartheta_i(t)$ in manifolds~\eqref{e19.03} and then in feedback control.

For load bus $i \in {\LL}$, we derive a similar internal model candidate by replacing $P_{M_i}(t)$ with $-P_{C_i}(t)$:
\begin{align}
\dot{\eta}_{i}(t) &= M_{i} \eta_{i}(t) - N_{i} (P_{C_i}(t)+P_{\NN_i}(t)).
\label{e26.2} 
\end{align}
For notional simplicity, we will use the augmented states $x_i(t) = [x_{i,1}(t),x_{i,2}(t),x_{i,3}(t),x_{i,4}^T(t)]^T=[w_i(t)$, $P_{M_i}(t)$, $P_{v_i}(t), \eta_i^T(t)]^T$ and manifolds $x_i^*(P_{L_i}(t),t) = [x_{i,1}^*(t)$, $x_{i,2}^*(t),x_{i,3}^*(t),(x_{i,4}^*(t))^T]^T= [w_i^*(t), P_{M_i}^*(P_{L_i}(t),t),$ $P_{v_i}^*(P_{L_i}(t),t),$ $\vartheta_i^T(t)]^T$ for $i \in {\GG}$ and use the augmented states $x_i(t) = [x_{i,1}(t),x_{i,2}(t),x_{i,4}^T(t)]^T = [w_i(t), P_{C_i}(t), \eta_i^T(t)]^T$ and manifolds $x_i^*(P_{L_i}(t),t) =[x_{i,1}^*(t),x_{i,2}^*(t),(x_{i,4}^*(t))^T]^T= [w_i^*(t), P_{C_i}^*(P_{L_i}(t),t), \vartheta_i^T(t)]^T$ for $i \in {\LL}$, where the dependency of $x_i^*$ on $P_{L_i}(t)$ is emphasized.

\subsection{Controller design}\label{sec:coo1}
We first conduct a coordinate transformation to convert the frequency control problem into a global stabilization
problem of the error dynamics with respect to manifolds~\eqref{e19.03} and~\eqref{e19.04}.
We make use of its unique lower triangular structure and apply a backstepping technique~\cite{KKK:95} to stabilize the error dynamics from the outer state to the inner state progressively.

Since internal model states $\eta_{i}(t)$ are expected to track $\vartheta_{i}(t)$ asymptotically for $\forall i \in {\VV}$, the estimation errors $\|\Psi_{i} T_{i}^{-1}(\rho_i)\eta_{i}(t)-\Psi_{i} T_{i}^{-1}(\rho_i)\vartheta_{i}(t)\|$ are expected to diminish.
By the certainty equivalent principle, we use the known term $\Psi_{i} T_{i}^{-1}(\rho_i)\eta_{i}(t)$
to replace unknown $P_{L_i}(t)=\Psi_{i} T_{i}^{-1}(\rho_i)\vartheta_{i}(t)$ when constructing the error dynamics.

Let us defined the tracking errors as follows: \begin{align} \tilde{x}_i(t)&\triangleq x_i(t)-x_i^*(\Psi_{i}T_i^{-1}(\rho_i)\eta_i(t),t) \label{e11_f0} \end{align} for $\forall i \in \VV$. Error dynamics for $i \in \GG$ become \begin{align} \dot{\tilde{x}}_{i,1}(t) &=  -\frac{1}{m_i}(D_i\tilde{x}_{i,1}(t)-\Psi_iT_i^{-1}(\rho_i)\tilde{x}_{i,4}(t)-\tilde{x}_{i,2}(t))\nnum\\ \dot{\tilde{x}}_{i,2}(t) &=  -\frac{1}{T_{CH_i}}(\tilde{x}_{i,2}(t)-K_{m_i}\tilde{x}_{i,3}(t))\nnum\\ \dot{\tilde{x}}_{i,3}(t) &=  -\frac{1}{T_{G_i}}(\tilde{x}_{i,3}(t)+\frac{K_{e_i}}{R_i}\tilde{x}_{i,1}(t)-\tilde{P}_{ref_i}(t))\nnum\\ &+\sum_{j \in \NN_i}t_{ij}(\Psi_iT_i^{-1}(\rho_j)\tilde{x}_{i,4}(t)-\Psi_jT_j^{-1}(\rho_j)\tilde{x}_{j,4}(t))\nnum\\ \dot{\tilde{x}}_{i,4}(t) &= \Phi_i(\rho_i)\tilde{x}_{i,4}(t) + N_i \tilde{x}_{i,3}(t) \label{e11_f1} \end{align} where $\tilde{P}_{ref_i}(t) \triangleq P_{ref_i}(t)-P_{ref_i}^*(\Psi_{i}T_i^{-1}(\rho_i)x_{i,4}(t),t)$. All the eigenvalues of $\Phi_i(\rho_i)$ are on the imaginary axis. Coordinate transformation $\hat{x}_{i,4}(t) \triangleq \tilde{x}_{i,4}(t) - \hat{x}_{i,4}^*(t)$ leads to \begin{align*} \dot{\hat{x}}_{i,4}(t) = M_i\hat{x}_{i,4}(t) + (m_iM_i+D_i I_{2\ell_i})N_i\tilde{x}_{i,1}(t) \end{align*} where $\hat{x}_{i,4}^*(t) = m_iN_i x_{i,1}(t)$. Since matrix $M_i$ is Hurwitz, the subsystem $\hat{x}_{i,4}(t)$ is input-to-state stable (ISS) regarding $\tilde{x}_{i,1}(t)$ as an external input.

Consider subsystem $\tilde{x}_{i,l-1}(t)$ in~\eqref{e11_f1} for $l=2,3$ and regard $\tilde{x}_{i,l}(t)$ as an external input. Tracking error $\tilde{x}_{i,l}(t)$ is designed to stabilize $\tilde{x}_{i,l-1}(t)$; i.e., the manifold $\hat{x}_{i,l}^*(t)$ of $\tilde{x}_{i,l}(t)$ cancels all the measurable terms in the dynamics of $\tilde{x}_{i,l-1}(t)$ and stabilizes it via $-k_{i,l-1}\tilde{x}_{i,l-1}(t)$. Apply the same idea to $i \in \LL$, then we have  \begin{align} \hat{x}_i(t) \triangleq  \tilde{x}_i(t)-\hat{x}_i^{*}(t) \label{e11} \end{align} and inputs \begin{align} &P_{ref_i}(t) = P_{ref_i}^*(\Psi_{i}T_i^{-1}(\rho_i)x_{i,4}(t),t)+(\frac{K_{e_i}}{R_i}+\frac{T_{CH_i}}{K_{m_i}}T_{G_i}\nnum\\ &\times (e^*_i+k_{i,1})(e^*_i+k_{i,2})(D_i-m_i\Psi_iT_i^{-1}(\rho_i)N_i))x_{i,1}(t)\nnum\\ &-\frac{T_{CH_i}}{K_{m_i}}T_{G_i}(e^*_i+k_{i,1})(e^*_i+k_{i,2})(x_{i,2}(t)-x_{i,2}^*(t))\nnum\\ &+T_{G_i}(\frac{1}{K_{m_i}}-\frac{T_{CH_i}}{K_{m_i}}(e^*_i+k_{i,1}+k_{i,2}))(\dot{x}_{i,2}(t)-\dot{x}_{i,2}^*(t))\nnum\\ &+ (x_{i,3}(t)-x_{i,3}^*(t))-k_{i,3}T_{G_i}\hat{x}_{i,3}(t)\nnum\\ &\lbd_i(t)=\lbd_i^*(\Psi_{i}T_i^{-1}(\rho_i)x_{i,4}(t),t)+m_i (e^*_i+k_{i,1})(c_i+k_{i,1})\nnum\\ &\times x_{i,1}(t)+(c_i+e^*_i+k_{i,1}+k_{i,2})\hat{x}_{i,2}(t) \label{e_input} \end{align} where $e_i^* \triangleq \Psi_iT_i^{-1}(\rho_i)N_i-\frac{D_i}{m_i}$, $\hat{x}_{i,1}^*(t)=0$, $\hat{x}_{i,2}^*(t) = -m_i (e^*_i+k_{i,1}) x_{i,1}(t)$, and $\hat{x}_{i,3}^*(t)=-m_i\frac{T_{CH_i}}{K_{m_i}}(e^*_i+k_{i,1}) (e^*_i+k_{i,2}) x_{i,1}(t)- (\frac{1}{K_{m_i}}- \frac{T_{CH_i}}{K_{m_i}}(e^*_i+k_{i,1}+k_{i,2}))(x_{i,2}(t)-x_{i,2}^*(t))$. Through transformations~\eqref{e11_f0},~\eqref{e11} and input~\eqref{e_input}, the augmented system, including~\eqref{e6.04},~\eqref{e24},~\eqref{eq_load},~\eqref{e26} and~\eqref{e26.2}, becomes
\begin{align}
\dot{\hat{x}}_{i}(t)&=A_i\hat{x}_{i}(t) + \sum_{j \in {{\NN}}_i}B_{ij}\hat{x}_{j,4}(t)
\label{e_result_1}
\end{align}
where
\begin{align}
&A_i =
\left[
\begin{array}{cccc}
-k_{i,1}&1/m_i&0&\frac{\Psi_i T_i^{-1}(\rho_i)}{m_i}\\
0&-k_{i,2}&{K_{m_i}}/{T_{CH_i}}&A_i(2,4)\\
0&0&-k_{i,3}&A_i(3,4)\\
A_i(4,1)&\bold{0}_{2\ell_i \times 1}&\bold{0}_{2\ell_i \times 1}&M_i\\
\end{array}
\right], i \in {\GG}\nnum\\
&A_i =
\left[
\begin{array}{ccc}
-k_{i,1}&1/m_i&\Psi_i T_i^{-1}(\rho_i)/m_i\\
0&-k_{i,2}&A_i(2,4)\\
A_i(4,1)&\bold{0}_{2\ell_i \times 1}&M_i\\
\end{array}
\right], i\in{\LL}
\label{e:matA0}
\end{align}
$A_i(2,4)=-(e_i^*+k_{i,1})\Psi_iT_i^{-1}(\rho_i)$,
$A_i(4,1)=(m_iM_i+$ $ D_i I_{2\ell_i})N_i$,
$A_i(3,4)=-\frac{1}{K_{m_i}}(T_{CH_i}(e_i^*+k_{i,1})(e^*_i+k_{i,2})$ $+\sum_{j \in {{\NN}}_i}t_{ij}) \Psi_iT_i^{-1}(\rho_i)$,
and $B_{ij}=[\bold{0}_{1 \times 2\ell_i}^T,\bold{0}_{1 \times 2\ell_i}^T,$ $-\frac{t_{ij}}{K_{m_i}}(\Psi_j T_j^{-1}(\rho_j))^T,$ $\bold{0}_{2 \ell_i \times 2 \ell_i}^T]^T$ for $i \in {\GG}$, $B_{ij}=\bold{0}_{(2 \ell_i+1) \times 1}$ for $i \in {\LL}$.
By the backstepping technique, the $k_i$ submatrix in~\eqref{e:matA0} is an upper-triangular Hurwitz matrix and $M_i$ is Hurwitz. This property is crucial for the stability of system~\eqref{e_result_1}.

The network-wide system becomes $\dot{\hat{x}}(t)=A\hat{x}(t)$ where $\hat{x}(t)=[\hat{x}_{1}^T(t),\cdots,\hat{x}_{N}^T(t)]^T$.
Since $\hat{x}_i^{*}(t)$ in~\eqref{e11} does not change the origin, the exponential stability of $\hat{x}_i(t)$ implies that the original system states $x_i(t)$, and inputs $u_i(t)$ in~\eqref{e6.04},~\eqref{e24},~\eqref{eq_load},~\eqref{e26} and~\eqref{e26.2} exponentially track their manifolds~\eqref{e19.03},~\eqref{e19.04} and $\vartheta_i(t)$. 

\subsection{Frequency stability guarantee}\label{sec:stab1}
The following theorem summarizes the exponential stability of system states $x_{i}(t)$ under distributed internal model controller~\eqref{e26},~\eqref{e26.2} and~\eqref{e_input} with respect to their manifolds~\eqref{e19.03},~\eqref{e19.04} and $\vartheta_i(t)$.
\begin{theorem}
Consider distributed control law~\eqref{e26},~\eqref{e26.2} and~\eqref{e_input}. Under Assumptions~\ref{asmp3} and~\ref{atsm1}, system states $x(t)$ are exponentially stable with respect to their manifolds~\eqref{e19.03},~\eqref{e19.04} and $\vartheta_i(t)$ if matrix $A$ is Hurwitz. In addition, there always exists a set of matrices $M_i,N_i$ and gains $k_{i,1}, k_{i,2}, k_{i,3}$ such that matrix $A$ is Hurwitz.
\label{attheo1}
\end{theorem}

In the proof, we provide Algorithm~\ref{p_algo} to identify a set of gains and matrices in a distributed way such that $A$ is Hurwitz.

\section{Controller synthesis for robust adaptive frequency control}\label{sec:sol2}
In this section, we study the case where the frequencies $\rho_{i}$ in exosystem~\eqref{e21} are unknown.
Internal models~\eqref{e26} and~\eqref{e26.2} will be used, but $T_i^{-1}(\rho_i)$ in~\eqref{e111} is uncertain to control authority $i$ due to the unknown frequencies $\rho_i$.
To address the challenge, we propose a new distributed adaptive internal model controller.

Let us define the augmented state $x_i(t) = [x_{i,1}(t)$, $x_{i,2}(t),x_{i,3}^T(t)]^T=[w_i(t), P_{M_i}(t), \eta_i^T(t)]^T$ (or $x_i(t) = [x_{i,1}(t), x_{i,2}(t),x_{i,3}^T(t)]^T=[w_i(t), P_{C_i}(t), \eta_i^T(t)]^T$ for $i \in \LL$) and corresponding manifolds $x_i^*(t)$.



\subsection{Controller design}\label{sec:coo2}
Like Section~\ref{sec:coo1}, a coordinate transformation is conducted to convert the global control problem into a global stabilization problem of the error dynamics.
Also, a backstepping approach is applied to ensure the stability of the error dynamics.
Consider the transformation
\begin{align}
\hat{x}_i(t)&\triangleq x_i(t)-x_i^*(\Lambda_i(t)x_{i,2}(t),t)-\hat{x}_i^{*}(t)
\label{coo21}
\end{align}
where $\hat{x}_{i,1}^*(t) =0, \hat{x}_{i,2}^*(t) = m_iN_i x_{i,1}(t)$.
The first two terms in~\eqref{coo21} define the tracking error and the third term is introduced by a backstepping technique to stabilize the error dynamics as~\eqref{e11}.
Consider inputs
\begin{align}
P_{M_i} (t)&= P_{M_i}^*(\Lambda_i(t)x_{i,3}(t),t)-m_i (k_{i}-\frac{D_i}{m_i}) x_{i,1}(t)\nnum\\
P_{C_i}(t)&=P_{C_i}^{*}(\Lambda_i(t)x_{i,3}(t),t)+m_i (k_{i}-\frac{D_i}{m_i}) x_{i,1}(t).
\label{e_input2}
\end{align}
Under coordinate transformation~\eqref{coo21}, the frequency control problem is transformed to a stabilization problem for the same reason as~\eqref{e11}. The difference is that we use estimated vector $\Lambda_i(t)$ instead of true vector $\Lambda_i^*(\rho_i)=\Psi_iT_i^{-1}(\rho_i)$.
The origin of the error dynamics does not change by $\hat{x}_i^{*}(t)$.

Through coordinate transformation~\eqref{coo21} and input~\eqref{e_input2}, systems~\eqref{eq_load},~\eqref{e24_2} and internal model~\eqref{e26} become
\begin{align}
\dot{\hat{x}}_{i}(t)&=A_i(\Lambda_i^*(\rho_i))\hat{x}_{i}(t) + B_{i}(\hat{\Lambda}_i(t))x_{i,3}(t)
\label{ate270}
\end{align}
for $\forall i \in {\VV}$ where $\hat{\Lambda}_i(t) \triangleq \Lambda_i(t)-\Lambda_i^*(\rho_i)$ is the estimation error
and
\begin{align*}
&A_i(\Lambda_i^*(\rho_i))=
\left[
\begin{array}{cc}
-k_{i}+\Lambda_i^*(\rho_i)N_i&\frac{1}{m_i} \Lambda_i^*(\rho_i)\\
(m_iM_i+D_i I) N_i&M_i\\
\end{array}
\right],\nnum\\
&B_i(\hat{\Lambda}_i(t))=
\left[
\begin{array}{c}
\frac{\hat{\Lambda}_i(t)}{m_i}\\
\bold{0}_{2\ell_i \times 2\ell_i}\\
\end{array}
\right].
\end{align*}

\subsection{Projected parameter estimator}\label{sec:adaptive2}
The quantity $\Lambda_i(t)$ is an estimate of $\Lambda_i^*(\rho_i)$ and its update law is given by:
\begin{align}
&\dot{\Lambda}_{i}^T(t)= J_{i}(t)-(\|J_{i}(t)\|+\gamma_i) \nnum\\
& \times (sgn ({\Lambda}_i(t)-\|\frac{(\rho_{\max}^2+1) \ell_i + \|M_i\|_F}{\|N_i\|}\|\textbf{1}_{2\ell_i \times 1})/2\nnum\\
&+sgn ({\Lambda}_i(t)+\|\frac{(\rho_{\max}^2+1) \ell_i + \|M_i\|_F}{\|N_i\|}\|\textbf{1}_{2\ell_i \times 1})/2)
\label{e:adap}
\end{align}
where $J_{i}(t)=-\frac{\hat{x}_{i,1}(t)}{m_i}x_{i,3}(t)$ and $\gamma_i>0$ is an arbitrary constant.
The first term $J_i(t)$ in~\eqref{e:adap} is designed to cancel cross term $\hat{\Lambda}_i(t)\frac{\hat{x}_{i,1}(t)}{m_i}x_{i,3}(t)$ by $\hat{\Lambda}_i(t)\dot{\hat{\Lambda}}_i^T(t)$ in the Lyapunov analysis.
The additional terms in~\eqref{e:adap} speed up the convergence rate by restricting the parameter estimates within $\|\Lambda_i(t)\| \leq \sqrt{2\ell_i}((\rho_{\max}^2+1) \ell_i + \|M_i\|_F)/\|N_i\|$. 
The bound is shown in Claim B in the proof of Theorem~\ref{attheo2}.

\subsection{Frequency stability guarantee}\label{sec:stab2}
The following theorem summarizes the asymptotic convergence of system states $x_{i}(t)$ to their manifolds~\eqref{e19.03L} and $\vartheta_i(t)$.
Consider matrix
\begin{align}
&\bar{A}_i=
\left[
\begin{array}{ccc}
\bar{A}_i(1,1)&((m_iM_i+D_i I) N_i)^T/2\\
(m_iM_i+D_i I) N_i/2&(M_i+M_i^T)/2+2I_{2\ell_i \times 2\ell_i}\\
\end{array}
\right]\nnum\\
&\bar{A}_i(1,1)=-k_{i}+(\rho_{\max}^2+1) \ell_i + \|M_i\|_F\nnum\\
&\quad +((\rho_{\max}^2+1) \ell_i + \|M_i\|_F)^2/(4m_i^2\|N_i\|^2).
\label{eq:A2}
\end{align}

\begin{theorem}
Consider distributed control law~\eqref{e26},~\eqref{e26.2} and~\eqref{e_input2} and adaptive law~\eqref{e:adap}. Under Assumptions~\ref{asmp3} and~\ref{atsm3}, system states $x(t)$ are asymptotically convergent to their manifolds~\eqref{e19.03L} and $\vartheta_i(t)$, if matrix $\bar{A}_i$ is negative definite for $\forall i \in {\VV}$.
In addition, there always exists a set of matrices $M_i,N_i$ and gain $k_{i}$ such that matrix $\bar{A}_i$ is negative definite.
\label{attheo2}
\end{theorem}

In the proof, we provide Algorithm~\ref{atalgo1} to identify a set of control gain and matrices in a distributed way such that $\bar{A}_i$ is negative definite.

\begin{remark}
Simplified synchronous generator model~\eqref{e24_2} (as well as the simplified controllable load model) prevent potential problems where the adaptive law relies on unmeasurable values $\vartheta_i(t)$ and $T_i^{-1}(\rho_i)$ to eliminate cross terms.
\oprocend
\label{rem:sec5}
\end{remark}

\section{Analysis}
This section presents the proofs of Theorem~\ref{attheo1} and~\ref{attheo2}.
\subsection{Proof of Theorem~\ref{attheo1}}
Assume that $A$ is Hurwitz. Then, linear time invariance system $\dot{\hat{x}}(t)=A\hat{x}(t)$ is exponentially stable.
Since coordinate transformation $\hat{x}_i^*(t)$ in~\eqref{e11} does not change the origin, this further implies that $x(t)$ in~\eqref{e24} is exponentially stable with respect to their manifolds~\eqref{e19.03},~\eqref{e19.04} and $\vartheta_i(t)$. One can prove the necessity part by reversing the steps above.

Now we proceed to prove the existence of control gains and matrices by construction. 
Consider system~\eqref{e_result_1} where matrices and control gains are chosen by Algorithm~\ref{p_algo}. We will show that $A$ is Hurwitz by verifying that the system is exponentially stable. 
\begin{algorithm}[t] \caption{Distributed selection of control gains} \label{p_algo}
\begin{algorithmic}[1]
\FOR{$i \in {\VV}$}
\STATE Choose a controllable pair $(M_i,C_i)$ such that $M_i$ is Hurwitz and $\lbd_{\max}(\frac{M_i+M_i^T}{2})<-3.5-|{\NN}_i|/(|{\NN}_i|+2)^2$;
\STATE Choose $0<\alpha_i < \frac{2 (-\lbd_{\max}(\frac{M_i+M_i^T}{2})-3.5-\frac{|{\NN}_i|}{(|{\NN}_i|+2)^2})^{\frac{1}{2}}}{\|(m_iM_i+D_i I_{2\ell_i})C_i\|}$;
\STATE $N_i = \alpha_i C_i$;
\STATE Find the solution $T_i^{-1}(\rho_i)$ of Sylvester equation~\eqref{e111};
\ENDFOR
\FOR{$i \in {\VV}$}
\STATE Choose $k_{i,1},k_{i,2},k_{i,3}$ sequentially such that\\
$k_{i,1} > \|\Psi_iT_i^{-1}(\rho_i)\|^2/(4 m_i^2)+1/(4m_i^2)+1.5,$\\ $k_{i,2} > \frac{K_{m_i}^2}{4T_{CH_i}^2}+(e^*_i+k_{i,1})^2\|\Psi_iT_i^{-1}(\rho_i)\|^2/4+1.5,$\\ $k_{i,3} > T_{CH_i}^2(e^*_i+k_{i,1})^2(e^*_i+k_{i,2})^2\|\Psi_iT_i^{-1}(\rho_i)\|^2/(4K_{m_i}^2)+\sum_{j \in {{\NN}}_i}t_{ij}^2(|{{\NN}}_i|+2)^2(\|\Psi_iT_i^{-1}(\rho_i)\|^2+\|\Psi_jT_j^{-1}\|^2)/(4K_{m_i}^2)+1.5.$
\ENDFOR
\end{algorithmic}
\end{algorithm}
Consider Lyapunov function candidate
$V_{i}(t)=\frac{1}{2} \|\hat{x}_{i}(t)\|^2$ for $\forall i \in {\VV}$.
Since $\hat{x}_{i}^T(t) A_i \hat{x}_{i}(t) \in {\mathbb R}$, $\hat{x}_{i}^T(t) A_i \hat{x}_{i}(t)=(\hat{x}_{i}^T(t) A_i \hat{x}_{i}(t))^T$. Hence, the Lie derivative of Lyapunov function candidate along the trajectories of system~\eqref{e_result_1} becomes
\begin{align*}
\dot{V}_{i}(t)&= \hat{x}_{i}^T(t) A_i \hat{x}_{i}(t)+\sum_{j \in {{\NN}}_i}\hat{x}_{i}(t)B_{ij}\hat{x}_{j,4}(t)\nnum\\
&= \hat{x}_{i}^T(t) \frac{A_i+A_i^T}{2} \hat{x}_{i}(t)+\sum_{j \in {{\NN}}_i}\hat{x}_{i}(t)B_{ij}\hat{x}_{j,4}(t).
\end{align*}
Since $\hat{x}_{i}(t)B_{ij}\hat{x}_{j,4}(t) \leq \frac{\delta}{2} \|\hat{x}_{i,3}(t)\|^2 + \frac{\|B_{ij}\|^2}{2\delta}\|\hat{x}_{j,4}(t)\|^2$ with $\delta=\frac{t_{ij}}{K_{m_i}}(|{{\NN}}_i|+2)^2\|\Psi_jT_j^{-1}(\rho_j)\|/2$, we have
\begin{align*}
\dot{V}_{i}(t)&\leq \hat{x}_{i}^T(t) \bar{A}_i \hat{x}_{i}(t)+\sum_{j \in {{\NN}}_i}\hat{x}_{j}^T(t)\bar{B}_{ij}\hat{x}_{j}(t)
\end{align*}
where 
\begin{align*}
&\bar{A}_i =\frac{A_i+A_i^T}{2}+
\left[
\begin{array}{ccc}
\bold{0}_{2 \times 2}& \bold{0}_{2 \times 1}&\bold{0}_{2 \times 2\ell_i}\\
\bold{0}_{1 \times 2}&P_i(2,2)& \bold{0}_{1 \times 2\ell_i}\\
\bold{0}_{2\ell_i \times 2}&\bold{0}_{2\ell_i \times 1}&\bold{0}_{2\ell_i \times 2\ell_i}\\
\end{array}
\right]
\end{align*}
and $P_i(2,2)=\sum_{j \in {{\NN}}_i}\frac{t_{ij}^2}{K_{m_i}^2}(|{{\NN}}_i|+2)^2\|\Psi_jT_j^{-1}(\rho_j)\|^2/4$
for $i \in {\GG}$ and $\bar{A}_i =\frac{A_i+A_i^T}{2}$ for $i\in{\LL}$, and $\|\bar{B}_{ij}\| \leq\frac{1}{(|{{\NN}}_i|+2)^2}$.

\textbf{Claim A:} It holds that $\hat{x}_i^T(t)\bar{A}_i\hat{x}_i(t) \leq -0.5\|\hat{x}_i(t)\|^2$.
\begin{IEEEproof}
Since $\hat{x}_{i,l}^T(t)\bar{A}_i(l,p)\hat{x}_{i,p}(t) \leq \frac{\delta}{2} \|\hat{x}_{i,l}(t)\|^2 + \frac{\|\bar{A}_i(l,p)\|^2}{2\delta}\|\hat{x}_{i,p}(t)\|^2$ for any $\delta>0$ and $\hat{x}_{i,4}^T(t)M_i \hat{x}_{i,4}(t) = \hat{x}_{i,4}^T(t)(\frac{M_i+M_i^T}{2}) \hat{x}_{i,4}(t)\leq \lbd_{\max}(\frac{M_i+M_i^T}{2})\|\hat{x}_{i,4}(t)\|^2$ by the Rayleigh quotient~\cite{parlett1974rayleigh}, we have
\begin{align}
\hat{x}_{i}^T(t) \bar{A}_i \hat{x}_{i}(t) \leq \hat{x}_{i}^T(t) A_i' \hat{x}_{i}(t) 
\label{e50}
\end{align}
where $A_i' = \diag{A_i'(1,1),A_i'(2,2),A_i'(3,3),A_i'(4,4)}$ for $i \in {\GG}$ and $A_i' = \diag{A_i'(1,1),A_i'(2,2),A_i'(4,4)}$ for $i \in {\LL}$, and
\begin{align*}
&A_i'(1,1)=-k_{1,k} + \frac{\|\Psi_iT_i^{-1}(\rho_i)\|^2}{4 m_i^2}+\frac{1}{4m_i^2}+1\nnum\\
&A_i'(2,2)=-k_{2,k} + \frac{K_{m_i}^2}{4T_{CH_i}^2}+\frac{(e_i^*+k_{i,1})^2\|\Psi_iT_i^{-1}(\rho_i)\|^2}{4} +1\nnum\\
&A_i'(3,3)=-k_{3,k}+1\nnum\\
&\quad +  T_{CH_i}^2(e_i^*+k_{i,1})^2(e^*_i+k_{i,2})^2\|\Psi_iT_i^{-1}(\rho_i)\|^2/(4K_{m_i}^2)\nnum\\
&\quad +\sum_{j \in {{\NN}}_i}\frac{t_{ij}^2}{4K_{m_i}^2}(|{{\NN}}_i|+2)^2(\|\Psi_iT_i^{-1}(\rho_i)\|^2+\|\Psi_jT_j^{-1}(\rho_j)\|^2)\nnum\\
&A_i'(4,4)=(\lbd_{\max}(\frac{M_i+M_i^T}{2}) + \frac{\|(m_iM_i+ D_i I_{2\ell_i})N_i\|^2}{4}\nnum\\
&\quad \ \ +3+|{{\NN}}_i|/(|{{\NN}}_i|+2)^2)I_{2\ell_i}.
\end{align*}
Algorithm~\ref{p_algo} ensures $A_i'(l,l)<-0.5$, and thus diagonal matrix $A_i'$ satisfies $\lbd_{\max}(A_i')<-0.5$. Hence, by~\eqref{e50} and the Rayleigh quotient, $\hat{x}_{i}^T(t) \bar{A}_i \hat{x}_{i}(t) \leq \lbd_{\max}( A_i')\|\hat{x}_{i}(t)\|^2 \leq -0.5\|\hat{x}_{i}(t)\|^2$.
\end{IEEEproof}
By Claim A,
\begin{align*}
\dot{V}_{i}(t) &\leq -V_i(t)+ \sum_{j \in {{\NN}}_i} \frac{\|\hat{x}_j(t)\|^2}{(|{{\NN}}_i|+2)^2}.
\end{align*}
Consider $U_i(t) =\frac{1}{2}\|\hat{x}_{i}(t)\|^2$ and $\dot{U}_i(t)=-U_i(t)+ \sum_{j \in {{\NN}}_i} \frac{\|\hat{x}_j(t)\|^2}{(|{{\NN}}_i|+2)^2}$.
By the comparison lemma (Lemma 3.4~\cite{Khalil:02}), it holds that $V_i(t) \leq U_i(t)$
for $t\geq 0$ when $V_i(0)\leq U_i(0)$.
The general solution $U_i(t)$ of the linear differential equation satisfies
\begin{align*}
U_i(t) &= e^{-t} U_i(0)+ \sum_{j \in {{\NN}}_i}\int_{0}^t e^{-(t-\tau)} \frac{\|\hat{x}_j(\tau)\|^2}{(|{{\NN}}_i|+2)^2} d\tau\nnum\\
&\leq  e^{-t} U_i(0)+ \sum_{j \in {{\NN}}_i}\frac{\|\hat{x}_j\|_{[0,t]}^2}{(|{{\NN}}_i|+2)^2} \int_{0}^t e^{-(t-\tau)} d\tau \nnum\\
&\leq e^{-t} U_i(0) + \sum_{j \in {{\NN}}_i} \frac{\|\hat{x}_j\|_{[0,t]}^2}{(|{{\NN}}_i|+2)^2}(1-e^{-t}) \nnum\\
&\leq e^{-t} U_i(0) + \sum_{j \in {{\NN}}_i} \frac{\|\hat{x}_j\|_{[0,t]}^2}{(|{{\NN}}_i|+2)^2}.
\end{align*}
Given $V_i(0) =U_i(0)$, it follows from $V_i(t) \leq U_i(t)$ that
\begin{align*}
V_i(t) \leq e^{- t} V_i(0) + \sum_{j \in {{\NN}}_i} \frac{\|\hat{x}_j\|_{[0,t]}^2}{(|{{\NN}}_i|+2)^2}.
\end{align*}
By taking norm, Cauchy-schwarz inequality and square-root to the above inequality, we have
\begin{align}
&\|\hat{x}_i(t)\| \leq e^{-0.5 t} \|\hat{x}(0)\| + \sum_{j \in {{\NN}}_i} \frac{\|\hat{x}_j\|_{[0,t]}}{|{{\NN}}_i|+2}\nnum\\
&\leq \max \{ (|{{\NN}}_i|+1) e^{-0.5t} \| \hat{x}_{i}(0)\|, \frac{|{{\NN}}_i|+1}{|{{\NN}}_i|+2}\max_{j \in  {{\NN}}_i }\{\|\hat{x}_{j}\|_{[0,t]}\}\}.
\label{el1}
\end{align}
Inequality~\eqref{el1} implies that $\hat{x}_i(t)$ is input-to-state stable (ISS)~\cite{sontag1989smooth} with respect to each $\hat{x}_j(t)$ with a contractive linear gain.
By distributed constrained small-gain theorem~\ref{ap-the3}, $\hat{x}(t)$ is exponentially stable.
The exponential stability of $\hat{x}(t)$ implies that matrix $A$ is Hurwitz.
\oprocend

\subsection{Proof of Theorem~\ref{attheo2}}
Consider Lyapunov function candidate
\begin{align*} 
V(t)=\frac{1}{2} \|\hat{x}(t)\|^2+\frac{1}{2}\sum_{i \in {\VV}}\|\hat{\Lambda}_{i}(t)\|^2.
\end{align*}
The Lie derivative of Lyapunov function candidate along the trajectories of system~\eqref{ate270} becomes
\begin{align} 
\dot{V}(t)&=\hat{x}^T(t)A(\Lambda^*(\rho))\hat{x}(t)+\sum_{i \in {\VV}}\hat{\Lambda}_{i}(t)(\dot{\hat{\Lambda}}_{i}^T(t)-J_{i}(t))
\label{ate20}
\end{align}
where $A(\Lambda^*(\rho)) = \diag{A_1(\Lambda_i^*(\rho_i)), \cdots,A_{|{\VV}|}(\Lambda_i^*(\rho_i))}$.
Claim B identifies an upper bound of uncertain term $\Lambda_i^*(\rho_i)$.

\textbf{Claim B}: 
$\|\Lambda_i^*(\rho_i)\| \leq ((\rho_{\max}^2+1) \ell_i + \|M_i\|_F)/\|N_i\|$.
\begin{IEEEproof}
By post-multiplying $T_i^{-1}(\rho_i)$ and taking Frobenius norm on both sides of Sylvester equation~\eqref{e111}, we have
\begin{align}
&\|N_i \Lambda_i^*(\rho_i)\|_F \leq \|T_i(\rho_i) \Phi_i(\rho_i) T_i^{-1}(\rho_i)\|_F +\|M_i\|_F.
\label{cm01}
\end{align}
By the definition of Frobenius norm, it holds that
\begin{align}
&\|N_i \Lambda_i^*(\rho_i)\|_F=\sqrt{\sum_{p=1}^{2\ell_i}\sum_{l=1}^{2\ell_i} N_{i,p}^2(\Lambda_{i,l}^*(\rho_i))^2}\nnum\\
&=\sqrt{(\sum_{p=1}^{2\ell_i} N_{i,p}^2)(\sum_{l=1}^{2\ell_i} (\Lambda_{i,l}^*(\rho_i))^2)}=\|N_i\|_F\| \Lambda_i^*(\rho_i) \|_{F}.
\label{ate28}
\end{align}
By~\eqref{ate28} and $\|\cdot\|_{F} \leq \|\cdot\|_{tr}$ (Lemma 10 in~\cite{srebro2004learning}),~\eqref{cm01} becomes
\begin{align*}
&\|N_i\|_F\|\Lambda_i^*(\rho_i)\|_F \leq \|T_i(\rho_i) \Phi_i(\rho_i) T_i^{-1}(\rho_i)\|_{tr} +\|M_i\|_F \nnum\\
&= \|\Phi_i(\rho_i) \|_{tr} +\|M_i\|_F \leq (\rho_{\max}^2+1) \ell_i + \|M_i\|_F.
\end{align*}
Note that $\|\cdot\|_F=\|\cdot\|_2$ for a vector.
\end{IEEEproof}

Claim C shows that $\hat{\Lambda}_{i}(t)(\dot{\hat{\Lambda}}_{i}^T(t)-J_{i}(t))$ is non-positive.

\textbf{Claim C}: $\hat{\Lambda}_{i}(t)(\dot{\hat{\Lambda}}_{i}^T(t)-J_{i}(t))\leq 0$.
\begin{IEEEproof}
If $|\hat{\Lambda}_{i,l}(t)| < \|\frac{(\rho_{\max}^2+1) \ell_i + \|M_i\|_F}{\|N_i\|}\|$ for all $l$, then $\hat{\Lambda}_{i}(t)(\dot{\hat{\Lambda}}_{i}^T(t)-J_{i}(t))= 0$. Assume there exists $l$ such that $|\hat{\Lambda}_{i,l}(t)| \geq \|\frac{(\rho_{\max}^2+1) \ell_i + \|M_i\|_F}{\|N_i\|}\|$.
Let $S^{+}_{i,l}(t)$ denote a set of indices $l$ such that $\hat{\Lambda}_{i,l}(t)\geq \|\frac{(\rho_{\max}^2+1) \ell_i + \|M_i\|_F}{\|N_i\|}\|$. Likewise, $S^{-}_{i,l}(t) \triangleq \{l|\hat{\Lambda}_{i,l}(t)\leq -\|\frac{(\rho_{\max}^2+1) \ell_i + \|M_i\|_F}{\|N_i\|}\|\}$.
Then, we have
\begin{align*}
&\hat{\Lambda}_{i}(t)(\dot{\hat{\Lambda}}_{i}^T(t)-J_{i}(t))\nnum\\
&\leq (\sum_{l \in S^{+}_{i,l}(t)}-\hat{\Lambda}_{i,l}(t)(\|J_{i}(t)\|+\gamma_i) \nnum\\
&+\sum_{l \in S^{-}_{i,l}(t)}\hat{\Lambda}_{i,l}(t)(\|J_{i}(t)\|+\gamma_i))\nnum\\
&\leq -\gamma_i(\sum_{l \in S^{+}_{i,l}(t)}|\hat{\Lambda}_{i,l}(t)| +\sum_{l \in S^{-}_{i,l}(t)}|\hat{\Lambda}_{i,l}(t)|)\leq0
\end{align*}
where $\hat{\Lambda}_{i,l}(t)=|\hat{\Lambda}_{i,l}(t)|$ for $l \in S^{+}_{i,l}(t)$ and $\hat{\Lambda}_{i,l}(t)=-|\hat{\Lambda}_{i,l}(t)|$ for $l \in S^{-}_{i,l}(t)$ are applied.
\end{IEEEproof}

By Claim B, we have 
\begin{align*}
&\|\hat{x}_{i,1}^T(t) \Lambda_i^*(\rho_i) \hat{x}_{i,3}(t)\| \leq \|\hat{x}_{i,1}^T(t)\| \|\Lambda_i^*(\rho_i)\hat{x}_{i,3}(t)\| \nnum\\
&\leq \frac{(\rho_{\max}^2+1) \ell_i + \|M_i\|_F}{\|N_i\|}(\frac{\delta\|\hat{x}_{i,1}(t)\|^2}{2} + \frac{\|\hat{x}_{i,3}(t)\|^2}{2\delta})
\end{align*}
and then
\begin{align} 
&\hat{x}^T(t)A(\Lambda^*(\rho))\hat{x}(t)\leq \hat{x}^T(t)\bar{A}\hat{x}(t)
\label{e:p200}
\end{align}
Symmetric matrix $\bar{A} = \diag{\bar{A}_1,\cdots,\bar{A}_{|{\VV}|}}$ is negative definite where $\bar{A}_i$ is defined in~\eqref{eq:A2}.
Thus, Claim C and~\eqref{e:p200} lead~\eqref{ate20} to 
\begin{align} 
\dot{V}(t)&\leq\hat{x}^T(t)\bar{A}\hat{x}(t) \leq \lbd_{\max}(\bar{A})\|\hat{x}(t)\|^2.
\label{e:p202}
\end{align}
Take the integral from $0$ to $t$ on both sides of~\eqref{e:p202}, then
\begin{align*}
- \lambda_{\max}(\bar{A}) \int_{0}^{t}\|\hat{x}(\tau)\|^2d\tau &\leq -\int_{0}^{t}\dot{V}(\tau)d\tau \leq V(0)<\infty
\end{align*}
where $V(t)\geq 0$ is applied.
Since $\int_{0}^{t}\|\hat{x}(\tau)\|^2d\tau$ is non-decreasing and upper bounded by $-V(0)/\lambda_{\max}(\bar{A})$, the limit $\lim_{t \rightarrow \infty}\int_{0}^{t}\|\hat{x}(\tau)\|^2d\tau$ exists.
Moreover, $\|\hat{x}(t)\|^2$ is uniformly continuous as shown in Claim D.

\textbf{Claim D}: $\|\hat{x}(t)\|^2$ is uniformly continuous.
\begin{IEEEproof}
Recall $\dot{V}(t)$ is non-positive. There exists a constant $U>0$ such that $\|\hat{x}(t)\| \leq U$ for $t \in [0,\infty)$. Consider
\begin{align}
&|\|\hat{{x}}(t+s)\|^2-\|\hat{{x}}(t)\|^2| \nnum\\
&= \sum_{i \in {\VV}}|\sum_{l=1}^3(\|\hat{x}_{i,l}(t+s)\|^2-\|\hat{x}_{i,l}(t)\|^2)|\nnum\\
&\leq  \sum_{i \in {\VV}}|\sum_{l=1}^3|\|\hat{x}_{i,l}(t+s)\|^2-\|\hat{x}_{i,l}(t)\|^2| \ |.
\label{ate341}
\end{align}
The term $\hat{x}_{i,l}(t+s)$ is given by $\hat{x}_{i,l}(t+s)  = \hat{x}_{i,l}(t) + \int_{t}^{t+s}\dot{\hat{x}}_{i,l}(\tau) d\tau$. By uniform boundedness of all $\hat{x}_{i,l}(t)$, for $l=1,2$ and any $s>0$, we have
\begin{align*}
\hat{x}_{i,l}(t) - a_{i,l} s \leq \hat{x}_{i,l}(t+s) \leq \hat{x}_{i,l}(t) + a_{i,l} s
\end{align*}
where $a_{i,1}$ is a positive constant and $a_{i,2}=a [1, \cdots, 1]^T$ is a vector with a positive constant $a$.
Therefore,
\begin{align*}
|\|\hat{x}_{i,l}(t+s)\|^2 -\|\hat{x}_{i,l}(t)\|^2| &\leq \|2 a_{i,l}^T \hat{x}_{i,l}(t) s\| + \| a_{i,l}^T a_{i,l} s^2\| \nnum\\
&\leq \|2 a_{i,l} s\|U + \| a_{i,l}^T a_{i,l} s^2\|
\end{align*}
where the right hand side is strictly increasing in $s$ and
$\lim_{s \rightarrow 0}\|2 a_{i,l} s\|U + \| a_{i,l}^T a_{i,l} s^2\| = 0$.
By applying the above bound to~\eqref{ate341}, we can prove the uniform continuity of $\|{x}(t)\|^2$; i.e., for any $\epsilon>0$, there always exists $\delta>0$ such that for all $t$ and $0 \leq s \leq \delta$, $|\|{x}(t+s)\|^2 -\|{x}(t)\|^2| \leq \epsilon$.
\end{IEEEproof}
It has been shown that $\|\hat{x}(t)\|^2$ is uniformly continuous, and 
$\lim_{t \rightarrow \infty}\int_{0}^{t}\|\hat{x}(\tau)\|^2d\tau$ exists and is finite. By the Barbalat's lemma (Lemma 8.2 in~\cite{Khalil:02}), $\|\hat{x}(t)\|^2$ asymptotically converges to zero.

\begin{algorithm}[t] \caption{Distributed selection of control gains}
\begin{algorithmic}[1]
\FOR{$i \in {\VV}$}
\STATE Choose a controllable pair $(M_i,C_i)$ such that $M_i$ is Hurwitz and $\lbd_{\max}(\frac{M_i+M_i^T}{2})<-1$;
\STATE Choose $0<\alpha_i < \frac{2 (-\lbd_{\max}(\frac{M_i+M_i^T}{2})-1)^{\frac{1}{2}}}{\|(m_iM_i+D_i I_{2\ell_i})C_i\|}$;
\STATE $N_i = \alpha_i C_i$;
\STATE Choose $k_{i}$ such that $k_{i}>(\rho_{\max}^2+1) \ell_i +\frac{((\rho_{\max}^2+1) \ell_i + \|M_i\|_F)^2}{4 m_i^2 \|N_i\|^2 } +1+ \|M_i\|_F$.
\ENDFOR
\end{algorithmic}
\label{atalgo1}
\end{algorithm}
Now we proceed to prove the existence of matrices and control gain such that matrix $\bar{A}_i$ is negative definite by construction.
Consider a set of matrices and control gain by Algorithm~\ref{atalgo1}.
Since $\hat{x}_{i,l}^T(t)\bar{A}_i(l,p)\hat{x}_{i,p}(t) \leq \frac{\delta}{2} \|\hat{x}_{i,l}(t)\|^2 + \frac{\|\bar{A}_i(l,p)\|^2}{2\delta}\|\hat{x}_{i,p}(t)\|^2$ for any $\delta>0$ and $\hat{x}_{i,2}^T(t)M_i\hat{x}_{i,2}(t)=\hat{x}_{i,2}^T(t)\frac{M_i+M_i^T}{2}\hat{x}_{i,2}(t)\leq \lbd_{\max}(\frac{M_i+M_i^T}{2})\|\hat{x}_{i,2}(t)\|^2$, we have 
\begin{align*}
\hat{x}_{i}^T(t) \bar{A}_i \hat{x}_{i}(t) \leq \hat{x}_{i}^T(t) A_i' \hat{x}_{i}(t) 
\end{align*}
where $A_i' = \diag{A_i'(1,1),A_i'(2,2)}$ and
$A_i'(1,1)=-k_{i} + (\rho_{\max}^2+1) \ell_i + \|M_i\|_F +1
+\frac{((\rho_{\max}^2+1) \ell_i + \|M_i\|_F)^2}{4m_i^2\|N_i\|^2}$,
$A_i'(2,2)=(\lbd_{\max}(\frac{M_i+M_i^T}{2}) +\frac{\|(m_iM_i+I D_i)N_i\|^2}{4}+1)I_{2 \ell_i}$.
Algorithm~\ref{atalgo1} ensures $A_i'(l,l)<0$, and thus diagonal matrix $A_i'$ is negative definite. This implies that $\bar{A}_i$ is also negative definite.
\oprocend

\section{Simulation}\label{sec:sim}
In this section, we present simulations to show the performance of the proposed distributed controllers.
All of the parameters are adopted from~\cite{Alvarado.Meng.ea:01,Kundar.Balu.Lauby:94}.
Consider the single line diagram of the IEEE 68-bus test system topology shown in~\cite{pal2006robust,rogers2012power}.
The network includes 16 generators ($|{\GG}|=16$) and 52 load buses ($|{\LL}|=52$). We assume that each generator/load bus $i \in {\GG}, {\LL}$ has (unknown) local net load $P_{L_i}(t) = 0.05 \sin (0.1 t) + 0.05 \sin (0.2 t)$ with $\Psi_{i} = [1, 0, 1, 0]$. Frequency upper bound is given by $\rho_{\max}=0.9$.


\textbf{System and controller parameters}. 
The generator parameters are adopted from page 598 in~\cite{Kundar.Balu.Lauby:94}:
$R_i = 0.05, \
T_{CH_i} = 0.3$, $T_{G_i} = 0.2$, $K_{m_i}=1$ and $K_{e_i}=1$ for $\forall i \in {\GG}$, and $m_i = 10$, $D_i = 1$ and $t_{ij} = 1.5 $ for $\forall i \in {\VV}$.
Demand response parameters $b_i=(40 \$/MWh)/(150s)$ and $c_i=(-0.8 \$/MW^2h)/(150s)$ for $i \in \LL$ are borrowed from~\cite{Alvarado.Meng.ea:01}. 

For the \emph{robust frequency control},
we choose $k_i=[1,26,99]^T$ and matrices
\begin{align*}
M_{i}=
\left[
\begin{array}{cccc}
-5.9&-2.1&-0.1&1.5\\
2.4&-6.3&-0.2&2.9\\
0.8&0.9&-6.6&2.5\\
1.6&0.3&0.8&-7\\
\end{array}
\right], \ 
N_{i}=
\left[
\begin{array}{c}
0.11\\
-0.1\\
0.12\\
0.12\\
\end{array}
\right]
\end{align*}
which satisfy that matrix $A$ is Hurwitz in Theorem~\ref{attheo1}.
For the \emph{robust adaptive frequency control} problem,
we choose $k_i=45.5$ and the above matrices, which guarantee that $\bar{A}_i$ is negative definite in Theorem~\ref{attheo2}.

\textbf{Simulation Results.}
\begin{figure}[t]   \centering   \includegraphics[width = \linewidth]{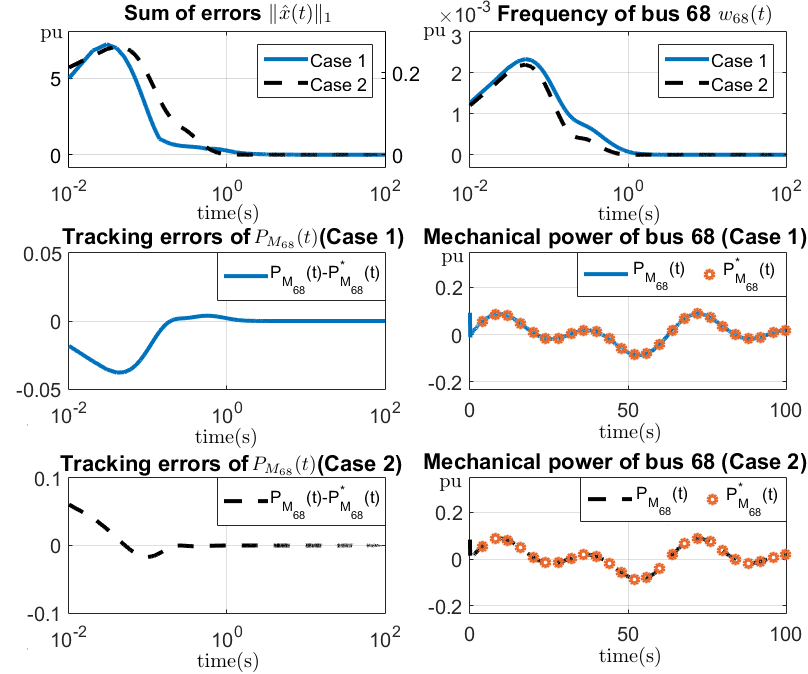}   \caption{
Simulation results (the solid lines for Case1, and the dashed lines for Case 2).}
\label{Combined_figure} \end{figure}
The solid lines in Figure~\ref{Combined_figure} summarize the results for the \emph{robust frequency control}.
In each subfigure, the horizontal axis represents time in log-scale or linear-scale, and the vertical axis represents corresponding values in per unit.
The first subfigure shows that the total state errors $\|\hat{x}(t)\|_1$ are exponentially stable (Y-axis on the left is for Case 1); i.e., the designed distributed controller achieves the objective and eventually steers network-wide frequency deviations to $0$. 
This implies that the signals of controllable load $P_{C_i}(t)$ of bus $i \in \LL$ and mechanical power $P_{M_{i}}(t)$ for $i \in {\GG}$ track desired manifolds that reject uncertain net loads; e.g., mechanical power $P_{M_{68}}(t)$ for $68 \in {\GG}$ rejects net loads as shown in the third and fourth subfigures.
Moreover, the frequency errors $\Delta w_{i}(t)$ are stable; e.g., frequency of bus $68$ is shown in the second subfigure.

The dashed lines in Figure~\ref{Combined_figure} summarize the results for the \emph{robust adaptive frequency control}.
The first subfigure shows that network-wide state errors converge to zero (Y-axis on the right is for Case 2), implying that the angular frequencies are controlled to $60\textit{Hz}$; e.g., the second subfigure, and $P_{C_i}(t)$ and $P_{M_{i}}(t)$ balance the local power demand and generation (as well as incoming and outgoing power).

The transient performance of the both cases in Figure~\ref{Combined_figure} look similar to each other, while their theoretic guarantees in the theorems are different.
The theoretic guarantees are valid for the worst case. That is, no matter what system parameters are, the robust controller always ensures exponential stability and the robust adaptive controller always ensures asymptotic stability. However, there could be some instances where the robust adaptive controller performs as good as or even better than the robust controller. The simulation in the article is actually one of these cases, and both controllers achieve exponential stability. These cases do not violate the theorems.

\section{Conclusion}
We have investigated the frequency control of multi-machine power systems subject to uncertain and dynamic net loads.
The proposed distributed internal model controllers coordinate synchronous generators and demand response to ensure frequency stability.
Simulations on the IEEE 68-bus test system demonstrate the performance of the controllers.

\appendix


\section{Appendix}
Distributed constrained small-gain theorem is introduced in this appendix.
The theorem is an extension of constrained small-gain theorem in~\cite{MZ-JH:CIS06} to a network set-up.

Consider an undirected graph $({\VV},\EE)$ and set ${{\NN}}_i \triangleq \{j \in {\VV} \setminus \{i\} | (i,j) \in \EE\}$.
The dynamic system associated with node $i$ is given by
\begin{align}
\dot{x}_i(t) = f_i(x(t),d_i(t),t)
\label{e:ap01}
\end{align}
where $x_i(t)$ and $d_i(t)$ denote system state and uncertainty respectively.
\begin{assumption}
The system~\eqref{e:ap01} is input-to-state stable with respect to neighboring states. Equivalently, there exist class $\mathcal{KL}$ function $\beta_i$ and class $\mathcal{K}$ functions $\gamma_{id}$ and $\gamma_{ij}$ such that for $\forall t \geq t_0$ and $\forall i \in {\VV}$,
\label{ap-asm1}
\begin{align}
\|x_i(t)\| &\leq \max\{\beta_i(\|x_i(t_0)\|, t-t_0), \gamma_{id}(\|d_i\|_{[t_0, t]}), \nnum\\
&\max_{j \in {{\NN}}_i}\{\gamma_{ij}(\|x_j\|_{[t_0,t]})\}\}.
\label{ap-e12}
\end{align}
\end{assumption}

\begin{assumption}
Gain functions $\gamma_{ij}$ are contraction mappings for $(i,j) \in \EE$; i.e., $\gamma_{ij}(s) < s$ for all $s>0$.
\label{ap-asm2}
\end{assumption}

\begin{theorem}\textbf{(Distributed constrained small-gain theorem)}
Under Assumptions~\ref{ap-asm1} and~\ref{ap-asm2}, the system~\eqref{e:ap01} is ISS with respect to $d$. Equivalently, there exists class $\mathcal{KL}$ function $\beta$ and class $\mathcal{K}$ function $\gamma_{id}$ such that for all $x_i(t_0) \in \hat{X}_i$ and $\|d\|_{[t_0,\infty)}<\hat{\Delta}_d$, the solution of~\eqref{e:ap01} exists and for $\forall t \geq t_0$,
\begin{align}
\|x(t)\| &\leq \max\{\beta(\|x(t_0)\|, t-t_0), \gamma_{id}(\|d(t)\|_{[t_0,t]}).
\label{theoremeq1}
\end{align}
Moreover, the function
$\beta(x,t) = |{\VV}|\sum_{i \in {\VV}}\beta_i(|{\VV}|\sum_{k \in {\VV}}$ $\beta_k(x,0),\frac{t}{(2\LL)^{|{\VV}|-1}})$
is a class $\KK\LL$ function candidate of $\beta(\cdot)$ in~\eqref{theoremeq1} where $\LL >1$ is a constant.
\label{ap-the3}
\end{theorem}
\begin{IEEEproof} For the notational simplicity in the sequent proof, we assume that ${\VV}$ is complete; i.e., ${{\NN}}_i = {\VV}\setminus\{i\}$. If $(i,j)\notin \EE$, then $\gamma_{ij}(s) = s$. We divide the remaining of the proof into three claims.

\textbf{Claim E:} The following hold for $i\in S_{\ell}\triangleq \{1,\cdots,\ell\}$: \begin{align}&\|x_i\|_{[t_0,T]}\leq \max\{\beta_i(\|x_i(t_0)\|,0),\nnum\\
&\max_{\substack{(i,i_1,\cdots,i_\kappa)\in\mathcal{P}_{i i_\kappa}\\
i_1,\cdots,i_\kappa\in S_{\ell}}}\gamma_{ii_1}\circ\cdots
\circ\gamma_{i_{\kappa-1}i_{\kappa}}\circ\gamma_{i_{\kappa}d}
(\|d_{i_{\kappa}}\|_{[t_0,T]}),\nnum\\
&\max_{j\in S_{\ell}\setminus\{i\}}\max_{\substack{(j,i_\kappa,\cdots,i)\in\mathcal{P}_{ji}\\
i_1,\cdots,i_{\kappa}\in S_{\ell}}}\nnum\\
&\gamma_{ii_1}\circ\gamma_{i_1i_2}\circ\cdots\circ\gamma_{i_\kappa j}\circ
\beta_j(\|x_j(t_0)\|,0),\nnum\\
&\max_{j\notin S_{\ell}}\max_{\substack{(i,i_1,\cdots,i_\kappa,j)\in\mathcal{P}_{ij}\\
i_1,\cdots,i_\kappa\in S_{\ell}}}\gamma_{ii_1}\circ\cdots\circ\gamma_{i_{\kappa}j}(\|x_j\|_{[t_0,T]})\}.
\label{ap-e14}\end{align}
\begin{IEEEproof} By~\eqref{ap-e12}, one can see that
\begin{align}\|x_1\|_{[t_0,T]}\leq &\max\{\beta_1(\|x_1(t_0)\|,0),\gamma_{1d}(\|d_1\|_{[t_0,T]}),\nnum\\
&\max_{j\neq 1}\{\gamma_{1j}(\|x_j\|_{[t_0,T]})\}\},\label{ap-e15}\end{align}
and \begin{align}\|x_2\|_{[t_0,T]}\leq &\max\{\beta_2(\|x_2(t_0)\|,0),\gamma_{2d}(\|d_2\|_{[t_0,T]}),\nnum\\
&\max_{j\neq 2}\{\gamma_{2j}(\|x_j\|_{[t_0,T]})\}\}.\label{ap-e16}\end{align}
Substitute~\eqref{ap-e16} into~\eqref{ap-e15}, and it renders the following:
\begin{align}\|x_1\|_{[t_0,T]}\leq &\max\{\beta_1(\|x_1(t_0)\|,0),\gamma_{1d}(\|d_1\|_{[t_0,T]}),\nnum\\
&\gamma_{12}\circ\beta_2(\|x_2(t_0)\|,0),\gamma_{12}
\circ\gamma_{2d}(\|d_2\|_{[t_0,T]}),\nnum\\
&\max_{j\neq 2}\{\gamma_{12}\circ\gamma_{2j}(\|x_j\|_{[t_0,T]})\},\nnum\\
&\max_{j\notin\{1,2\}}\{\gamma_{1j}(\|x_j\|_{[t_0,T]})\}\}.\label{ap-e22}\end{align}
Since $\gamma_{12}\circ\gamma_{21}$ is a contraction mapping, it follows from~\eqref{ap-e22} that \begin{align}\|x_1\|_{[t_0,T]}\leq &\max\{\beta_1(\|x_1(t_0)\|,0),\gamma_{1d}(\|d_1\|_{[t_0,T]}),\nnum\\
&\gamma_{12}\circ\beta_2(\|x_2(t_0)\|,0),\gamma_{12}
\circ\gamma_{2d}(\|d_2\|_{[t_0,T]}),\nnum\\
&\max_{j\notin\{1,2\}}\{\max\{\gamma_{1j},
\gamma_{12}\circ\gamma_{2j}\}(\|x_j\|_{[t_0,T]})\}\}.\label{ap-e17}\end{align}
By symmetry, one can show a similar property to~\eqref{ap-e17} for $\|x_2\|_{[t_0,T]}$. So~\eqref{ap-e14} holds for the case of $\ell=2$.
Now assume that~\eqref{ap-e14} holds for some $\ell < n$. Similar to~\eqref{ap-e15}, we have \begin{align}&\|x_{\ell+1}\|_{[t_0,T]}\leq \max\{\beta_{\ell+1}(\|x_{\ell+1}(t_0)\|,0),\nnum\\
&\gamma_{(\ell+1)d}(\|d_{\ell+1}\|_{[t_0,T]}),\max_{j\neq (\ell+1)}\{\gamma_{(\ell+1)j}(\|x_j\|_{[t_0,T]})\}\}.\label{ap-e18}\end{align}
Following analogous steps above, one can show that~\eqref{ap-e15} holds for $\ell+1$. By induction, we complete the proof.
\end{IEEEproof}

\textbf{Claim F:} The solution to~\eqref{e:ap01} exists and it is bounded.
\begin{IEEEproof} A direct result of Claim E is that the following holds for all $i\in {\VV}$:
\begin{align}&\|x_i\|_{[t_0,T]}\leq \max\{\beta_i(\|x_i(t_0)\|,0),\nnum\\
&\max_{\substack{(i,i_1,\cdots,i_\kappa)\in\mathcal{P}_{i i_\kappa}\\
i_1,\cdots,i_\kappa\in {\VV}}}\gamma_{ii_1}\circ\cdots
\circ\gamma_{i_{\kappa-1}i_{\kappa}}\circ\gamma_{i_{\kappa}d}
(\|d_{i_{\kappa}}\|_{[t_0,T]}),\nnum\\
&\max_{j\neq i}\max_{\substack{(j,i_\kappa,\cdots,i)\in\mathcal{P}_{ji}\\
i_1,\cdots,i_{\kappa}\in {\VV}}}\nnum\\
&\gamma_{ii_1}\circ\gamma_{i_1i_2}\circ\cdots\circ\gamma_{i_\kappa j}\circ
\beta_j(\|x_j(t_0)\|,0)\}.\label{ap-e19}\end{align}
Since all the gain functions $\gamma_{ij}$ are contraction mappings,~\eqref{ap-e19} renders the following:
\begin{align}
&\|x_i\|_{[t_0,T]}\leq \max\{\beta_i(\|x_i(t_0)\|,0),\nnum\\
&\max_{j\in{{\NN}}_i}\gamma_{ij}\circ\gamma_{jd}(\|d_j\|_{[t_0,T]}),\max_{j\in {{\NN}}_i}\beta_j(\|x_j(t_0)\|,0)\}.\label{ap-e20}\end{align}
Because of the choice of $x_i(t_0)$ and the bound on $d$, the relation~\eqref{ap-e19} holds for any $T$. It implies that
\begin{align*}
&\|x_i(t)\|\leq \max\{\beta_i(\|x_i(t_0)\|,0),\hat{\Delta}_d,\max_{j\in {{\NN}}_i}\beta_j(\|x_j(t_0)\|,0)\}.
\end{align*}
for all $t\geq t_0$ and thus is uniformly bounded. It completes the proof.
\end{IEEEproof}

\textbf{Claim G:}
System~\eqref{e:ap01} is ISS; i.e., the following holds for all $i\in S_\ell\triangleq \{1,\cdots,\ell\}$:
\begin{align}
\|x_i(t)\|&\leq \max\{\tilde{\beta}^{[\ell-1]}_i(\|x\|_{\infty},t-t_0),
\gamma_i^{[\ell-1]}(\|d\|_{[t_0,t]}),\nnum\\
&\max_{j\notin S_{\ell}}\max_{\substack{(i,i_1,\cdots,i_\kappa,j)\in\mathcal{P}_{ij}\\
i_1,\cdots,i_\kappa\in S_{\ell}}}\gamma_{ii_1}\circ\cdots\circ\gamma_{i_{\kappa}j}(\|x_j\|_{[t_0,t]})\},
\label{ap-e8}
\end{align}
for some class $\mathcal{KL}$ function $\tilde{\beta}^{[\ell-1]}_i$ where $\|x\|_{\infty} \triangleq \sup\{\|x(t)\| \ | \ t \in [t_0, \infty]\}$.
\begin{IEEEproof}
Let $\ell=2$. Note that for any constant $\LL>1$,
\begin{align}
&\|x_1(t_0+T)\|\leq \max\{\beta_1(\|x_1(t_0+\frac{2\LL-1}{2\LL}T)\|,\frac{1}{2\LL}T),\nnum\\
&\gamma_{1d}(\|d_1\|_{[t_0+\frac{2\LL-1}{2\LL}T,t_0+T]}),\max_{j\neq1}\gamma_{1j}(\|x_j\|_{[t_0+\frac{2\LL-1}{2\LL}T,t_0+T]})\}\nnum\\
&\leq \max\{\beta_1(\|x\|_{\infty},\frac{1}{2\LL}T),\gamma_{1d}(\|d_1\|_{[t_0+\frac{2\LL-1}{2\LL}T,t_0+T]}),\nnum\\
&\max_{j\neq1}\{\gamma_{1j}(\|x_j\|_{[t_0+\frac{2\LL-1}{2\LL}T,t_0+T]})\}\}.
\label{ap-e2}
\end{align}
For any $\tau_2\in[\frac{2\LL-1}{2\LL}T, T]$, it holds that \begin{align}&\|x_2(t_0+\tau_2)\|\nnum\\
&\leq \max\{\beta_2(\|x_2(t_0+\frac{2\LL-2}{2\LL}T)\|,\tau_2-\frac{2\LL-2}{2\LL}T),\nnum\\
&\gamma_{2d}(\|d_2\|_{[t_0+\frac{2\LL-2}{2\LL}T,t_0+\tau]}),\nnum\\
&\max_{j\neq2}\{\gamma_{2j}(\|x_j\|_{[t_0+\frac{2\LL-2}{2\LL}T,t_0+\tau]})\}\}\nnum\\
&\leq \max\{\beta_2(\|x\|_{\infty},\frac{1}{\LL}T),
\gamma_{2d}(\|d_2\|_{[t_0+\frac{2\LL-2}{2\LL}T,t_0+T]}),\nnum\\
&\max_{j\neq2}\{\gamma_{2j}(\|x_j\|_{[t_0+\frac{2\LL-2}{2\LL}T,t_0+T]})\}\}.
\label{ap-e3}\end{align}
So~\eqref{ap-e3} implies that \begin{align}\|x_2\|&_{[t_0+\frac{2\LL-1}{2\LL}T, t_0+T]}\leq\max\{\beta_2(\|x\|_{\infty},\frac{1}{\LL}T),\nnum\\
&\gamma_{2d}(\|d_2\|_{[t_0+\frac{2\LL-2}{2\LL}T,t_0+T]}),\nnum\\
&\max_{j\neq2}\{\gamma_{2j}(\|x_j\|_{[t_0+\frac{2\LL-2}{2\LL}T,t_0+T]})\}\}.
\label{ap-e4}\end{align}
Substitute~\eqref{ap-e4} into~\eqref{ap-e2}, and we have \begin{align}&\|x_1(t_0+T)\|\nnum\\
&\leq \max\{\beta_1(\|x\|_{\infty},\frac{1}{2\LL}T),
\gamma_{1d}(\|d_1\|_{[t_0+\frac{2\LL-2}{2\LL}T,t_0+T]}),\nnum\\
&\gamma_{12}\circ\beta_2(\|x\|_{\infty},\frac{1}{\LL}T),\gamma_{12}\circ\gamma_{2d}(\|d_2\|_{[t_0+\frac{2\LL-2}{2\LL}T,t_0+T]}),\nnum\\
&\gamma_{12}\circ\gamma_{21}(\|x_1\|_{[t_0+\frac{2\LL-2}{2\LL}T,t_0+T]}),\nnum\\
&\max_{j\notin S_2}\max\{\gamma_{1j},\gamma_{12}\circ\gamma_{2j}\}
(\|x_j\|_{[t_0+\frac{2\LL-2}{2\LL}T,t_0+T]})\}.\label{ap-e5}\end{align}
Since $\gamma_{12}\circ\gamma_{21}(\cdot)$ is a contraction mapping, there is class $\mathcal{KL}$ function $\tilde{\beta}_1$ such that \begin{align}\|x_1(t)\|&\leq \max\{\tilde{\beta}_1(\|x\|_{\infty},t-t_0),\nnum\\
&\gamma_{1d}(\|d_1\|_{[t_0,t]}),\gamma_{12}\circ\gamma_{2d}(\|d_2\|_{[t_0,t]}),\nnum\\
&\max_{j\notin S_2}\max\{\gamma_{1j},\gamma_{12}\circ\gamma_{2j}\}(\|x_j\|_{[t_0,t]})\}.
\label{ap-e6}\end{align}
By symmetry, there is class $\mathcal{KL}$ function $\tilde{\beta}_2$ such that \begin{align}\|x_2(t)\|&\leq \max\{\tilde{\beta}_2(\|x\|_{\infty},t-t_0),\nnum\\
&\gamma_{2d}(\|d_2\|_{[t_0,t]}),
\gamma_{21}\circ\gamma_{1d}(\|d_1\|_{[t_0,t]}),\nnum\\
&\max_{j\notin S_2}\max\{\gamma_{2j},\gamma_{21}\circ\gamma_{1j}\}
(\|x_j\|_{[t_0,t]})\}.\label{ap-e7}\end{align} Hence, we have shown that~\eqref{ap-e8} holds for $\ell=2$.
Now assume~\eqref{ap-e8} holds for some $\ell<n$. Recall that 
\begin{align}\|x&_{\ell+1}(t)\|\leq \max\{\beta_{\ell+1}(\|x_{\ell+1}(t_0)\|,t-t_0),\nnum\\
&\gamma_{\ell+1}(\|d_{\ell+1}\|_{[t_0,t]}),
\max_{j\neq \ell+1}\gamma_{ij}(\|x_j\|_{[t_0,t]})\}.\label{ap-e9}\end{align} By using similar arguments towards the case of $\ell=2$, one can show~\eqref{ap-e8} holds for $\ell+1$.
Now we proceed to find a relation between $\|x\|_{\infty}$ and $\|d\|_{\infty}$.
Because $\|x_i(t_0)\|\leq \|x(t_0)\|$, note that
\begin{align*}\|x_i\|_{\infty}\leq &\max\{\beta_i(\|x(t_0)\|,0),\gamma_{id}(\|d_i\|_{[t_0,t]}),\\
&\max_{j\neq i}\{\gamma_{ij}(\|x_j\|_{\infty})\}\}.
\end{align*}
Similar to~\eqref{ap-e8}, one can show by induction that there are class $\mathcal{K}$ functions $\rho_i$ and $\rho_{id}$ such that
\begin{align}
\|x_i\|_{\infty}\leq \max\{\rho_i(\|x(t_0)\|),\rho_{id}(\|d_i\|_{[t_0,t]})\}.
\label{ap-e10}
\end{align}
The combination of~\eqref{ap-e10} and~\eqref{ap-e8} achieves the desired result. \end{IEEEproof}

Now proceed with the proof that function $\beta(x,t) = |{\VV}|\sum_{i \in {\VV}}\beta_i(|{\VV}|\sum_{k \in {\VV}}\beta_k(x,0),\frac{t}{(2\LL)^{|{\VV}|-1}})$ is a candidate of class $\KK\LL$ function $\beta$ in~\eqref{theoremeq1}.
We first find candidates of functions $\tilde{\beta}_i^{[\ell-1]}$ in~\eqref{ap-e8} and $\rho_i$ in~\eqref{ap-e10} and then combine them together.
Note that by substituting~\eqref{ap-e4} into~\eqref{ap-e2}, we have equation~\eqref{ap-e5}.
Consider class $\KK\LL$ functions in equation~\eqref{ap-e5}:
\begin{align*}
&\|x_1(t_0+T)\|\nnum\\
&\leq \max\{\beta_1(\|x\|_{\infty},\frac{1}{2\LL}T),\gamma_{12}\circ\beta_2(\|x\|_{\infty},\frac{1}{\LL}T)\}\nnum\\
&\leq \max\{\beta_1(\|x\|_{\infty},\frac{1}{2\LL}T)+\beta_2(\|x\|_{\infty},\frac{1}{2\LL}T)\}.
\end{align*}
This implies that, in~\eqref{ap-e6}, $\tilde{\beta}_1(x,t) = \sum_{k=1}^{2}\beta_{k}(x,\frac{t}{2\LL})$ is a $\KK \LL$ function candidate.
Likewise, in~\eqref{ap-e8},
\begin{align}
\tilde{\beta}_i^{[\ell-1]}(x,t)=\sum_{k=1}^{\ell}\beta_k(x,\frac{t}{(2\LL)^{\ell-1}})
\label{cooc1}
\end{align}
is a $\KK \LL$ function candidate for $\forall i \in S_{\ell}$ because we conduct $\ell-1$ times of the substitutions.
In a similar way, one can show that, in~\eqref{ap-e10},
\begin{align}
\rho_i(x) = \sum_{k=1}^{\ell}\beta_k(x,0)
\label{cooc2}
\end{align}
is a class $\KK$ function candidate for $\forall i \in S_{\ell}$.
Now we proceed to find a relation between $\tilde{\beta}_i^{[\ell-1]}$ and $\rho_i$ when $S_{\ell}={\VV}$.
With equation~\eqref{ap-e10},
\begin{align}
\|x\|_{\infty} &\leq \sum_{i \in {\VV}}\|x_i\|_{\infty} \nnum\\
&\leq |{\VV}| \max_{i \in {\VV}}\{\rho_{i}(\|x(t_0)\|),\rho_{id}(\|d_i\|_{[t_0,t]})\}.
\label{coo2}
\end{align}
By combining~\eqref{ap-e8} and~\eqref{coo2}, 
\begin{align*}
\|x_i(t)\|&\leq \max\{\tilde{\beta}^{[|{\VV}|-1]}_i (|{\VV}|\max_{k \in {\VV}}\rho_k(\|x(t_0)\|),t-t_0),\nnum\\
&\ \ \ \ \ \ \ \ \ \gamma_i^{[\ell-1]}(\|d\|_{[t_0,t]})\}.
\end{align*}
This implies that 
\begin{align}
\beta(x,t) = |{\VV}|\max_{i \in {\VV}}\tilde{\beta}^{[|{\VV}|-1]}_i(|{\VV}|\max_{k \in {\VV}}\rho_k(x),t)
\label{cooc3}
\end{align}
is one of the class $\KK\LL$ function candidates.
By applying~\eqref{cooc1} and~\eqref{cooc2} to~\eqref{cooc3}, we have the result.
\end{IEEEproof}
\begin{remark}
If functions $\beta_i(\cdot)$ in~\eqref{ap-e12} for $\forall i \in {\VV}$ are $\beta_i(x,t) =a_i^{-p_i(t)}r_i(x)$, then $\beta(\cdot)$ in~\eqref{theoremeq1} is also in the same form: $\beta(x,t) =a^{-p(t)}r(x)$
where $a, a_i>0$ are constants, $p(t), p_i(t)$ are increasing functions without bound and $r(x), r_i(x)$ are class $\KK$ functions.
\oprocend
\label{remark:last}
\end{remark}
Remark~\ref{remark:last} indicates that if functions $\beta_i(\cdot)$ are exponential functions, then $\beta(\cdot)$ is also an exponential function.
 
\end{document}